# A Quantitative Analysis of Possible Futures of Autonomous Transport

C L Benson PhD[a,b,c*] , P D Sumanth MSc[a], A P Colling MSc[a]

[a] *Delft University of Technology, Netherlands*
[b] *United States Air Force, United States*
[c] *Massachusetts Institute of Technology, Cambridge, United States*
* Corresponding Author. Email: cbenson07@gmail.com

**Synopsis**

Autonomous ships (AS) used for cargo transport have gained a considerable amount of attention in recent years. They promise benefits such as reduced crew costs, increased safety and increased flexibility. This paper explores the effects of a faster increase in technological performance in maritime shipping achieved by leveraging fast-improving technological domains such as computer processors, and advanced energy storage. Based on historical improvement rates of several modes of transport (Cargo Ships, Air, Rail, Trucking) a simplified Markov-chain Monte-Carlo (MCMC) simulation of an intermodal transport model (IMTM) is used to explore the effects of differing technological improvement rates for AS. The results show that the annual improvement rates of traditional shipping (Ocean Cargo Ships = 2.6%, Air Cargo = 5.5%, Trucking = 0.6%, Rail = 1.9%, Inland Water Transport = 0.4%) improve at lower rates than technologies associated with automation such as Computer Processors (35.6%), Fuel Cells (14.7%) and Automotive Autonomous Hardware (27.9%). The IMTM simulations up to the year 2050 show that the introduction of any mode of autonomous transport will increase competition in lower cost shipping options, but is unlikely to significantly alter the overall distribution of transport mode costs. Secondly, if all forms of transport end up converting to autonomous systems, then the uncertainty surrounding the improvement rates yields a complex intermodal transport solution involving several options, all at a much lower cost over time. Ultimately, the research shows a need for more accurate measurement of current autonomous transport costs and how they are changing over time.

**Biographical Notes:**
Christopher Benson is an Active Duty Officer in the US Air Force with experience as an engineer with F-16s, Space Systems, and Autonomous Aerial Systems. He is currently working as an exchange scientist with the Royal Netherlands Defence Academy at the Technical University of Delft, focusing on autonomous maritime ship design.

Pranav Sumanth is a PhD Candidate at TU Delft, studying logistics networks.

Alina Colling is a PhD Candidate at TU Delft, studying the design of a new waterborne transport concept for an integration of a higher level of autonomy on vessels.

## 1. Introduction: Benefits of Autonomous Maritime Shipping

In recent years, the maritime world has been increasingly interested in exploring the benefits of autonomous maritime vessels for freight transportation. This has resulted in a number of exploratory projects including the AAWA autonomous shipping concept (Rolls-Royce Marine, 2016), the Yara Birkeland electric, autonomous ship (Skredderberget, 2018), a Japanese Trans-Pacific test (Cooper and Matsuda, 2017), the MUNIN research project (2016), a Chinese AS test location (Jennings, 2018), an autonomous military ship (Mizokami, 2018), the DIMECC 'One Sea' Consortium (Haikkola, 2017), and a start-up company retrofitting old ships to be autonomous (Constine, 2018).

Many of these efforts focus on the expected benefits of AS, including reduced operational costs, reduced manning, increased operational times, reduced fuel consumption, improved lifestyles for the seafarers, and increased maritime shipping capacity (Kobyliński, 2018), among others. Others have shown more scepticism toward the proposed benefits and have pointed out many challenges that have not yet been solved including legal (Karlis, 2018), commercial (Willumsen, 2018) and Technical (Kobyliński, 2018).

One additional benefit that could be brought about by autonomous shipping is an increased technological improvement rate in the maritime shipping industry. This higher improvement rate is facilitated by increased potential for upgrading via software (Greengard, 2015), rapidly improving energy technologies for propulsion (van Biert et al, 2016) and 'Big Data' driven learning to continuously improve efficiency of transport (Perera and Mo, 2016). This paper explores potential effects of this final benefit by comparing the improvement rates of traditional freight transportation technologies with estimates of their autonomous counterparts.

**2. Historical Improvement Rates of Cargo Transportation Modes**

Cargo transportation involves a number of competing and complementary transportation modes including ocean cargo ships, inland water transport vessels, rail, trucking, and air cargo transport. While each of these has different characteristics, all have improved considerably since their introduction. One of the most basic measures of performance for cargo transport is the operational cost per distance per weight, which fits the criteria of a suitable metric for long-term performance measurement (Benson and Magee, 2015) as is shown in Equation 1, where distance can be measured in kilometres, weight is measured in metric tons, and cost is measured in inflation adjusted dollars.

$$\frac{Distance * Weight}{Cost} \qquad (1)$$

Using a generalized version of Moore's law and exponential regression, it is possible to track this measure over time to create a technological improvement rate for each of the cargo transportation modes (Benson and Magee, 2015). Table 1 shows the the technological improvement rates for each mode of cargo transport for the measure above.

Table 1: Historical Exponential Improvement Rates for Cargo Transportation Modes

| Domain | Improvement Rate | Years | Source |
|---|---|---|---|
| Inland Water Transport | 0.4% | 2008-2018 | US Federal Reserve, 2018 |
| Trucking | 0.6% | 1947-2001 | Glaeser and Kohlase, 2004 |
| Rail | 1.9% | 1890-2001 | Glaeser and Kohlase, 2004 |
| Ocean | 2.1% | 1954-2004 | Hummels, 2007 |
| Air Cargo | 5.5% | 1954-2004 | Hummels, 2007 |

These improvement rates are the culmination of all of the factors that affect marginal operational costs of shipping such as technological change, policy decisions, natural resource prices and the changing distribution of transport products and geographies. While there have been efforts to separate out these different factors (Jacks et al, 2008; North, 1958), the overall change in operational costs is used in the interest of simplicity and transparency.

As can be seen in Table 1, the improvement rates for all modes of cargo transport range between 0 and 6%, which are relatively low compared to those of Integrated Circuits (Moore's Law) which is 36.3%, Solar PV energy generation (9.5%), Fuel Cells (14.4%), and even electrochemical batteries (7.0%) (Benson, 2014).

There is evidence that sharing technological developments with faster improving technological domains can result in a new domain that improves more rapidly. One example of this is telecommunications, which originally improved at around 14.3% for 'wired information transmission' and, after integration of digital technologies largely based on the domain of integrated circuits, now improves at 50.4% in wireless information transmission (Benson and Magee, 2015). This brings us to the central question of this paper – how could a move into digital technologies that improve at a faster rate affect the cargo transportation industry?

## 3. Inter-Modal Transport Model (IMTM)

The future improvement rates of the cargo industry are impacted by a great number of factors including technology, shipping distances, cargo loads, types of cargo, tariffs, prices of fuel and other natural resources and the geography of the route. Due to these compounding factors there is significant uncertainty in the future performance, as measured by the metric in Equation 1, across the cargo transportation modes.

### 3.1. *Markov-Chain Monte-Carlo Structure of Intermodal Transport Model*

One way of handling this uncertainty is to build in uncertainty around the metrics through probability distributions and to run a large number of iterations, thus finding a range of probability for future outcomes. This was done by Parno (2015) as a simulation of transport maps using a Markov-Chain Monte-Carlo model. The authors use a similar model for the analysis of future performance change for multi-modal cargo transport, with a main difference being a focuses on how the distribution of transport performance (i.e. costs) change over time between the different competing modes of transport. Figure 1 shows the steps in the simplified Intermodal Transport Model as defined by a Markov-Chain Monte-Carlo model and is followed by a step-by-step description of the model below.

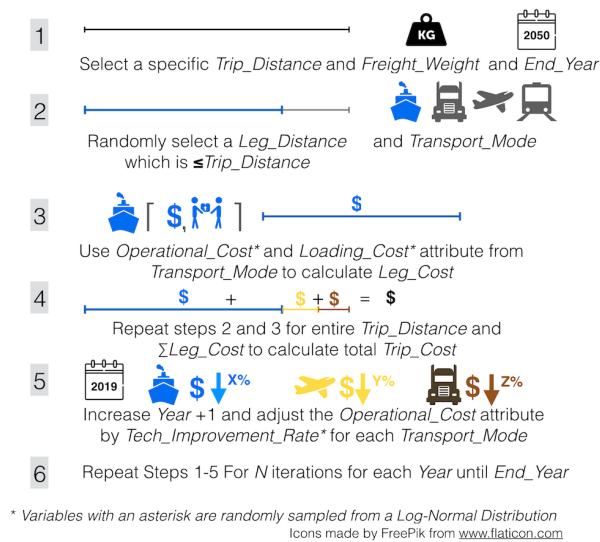

Figure 1: Summary of Markov-Chain Monte-Carlo Inter-Modal Transport Model

At a high level, the 'Markov-Chain' aspect of the model acts as a 'random-walk' through any shipping route, which gives no pre-conceived notions to which mode should be used next. The 'Monte-Carlo' aspect is captured by the fact that the attribute variables for each of the transport modes are represented as distributions and not as single values, and thus running many iterations will provide a general understanding of a large number of possible futures. The final result of this model is not a given prediction of what the future will look like, but rather reduced uncertainty around future scenarios.

In Step 1 of the model, the input parameters for the test are set, including the Trip_Distance (measured in kilometres) and Freight_Weight (measured in metric tonnes), and the End_Year, which is how far into the future the model will run (in our case, until 2050).

Step 2 begins the Markov-Chain, whereby the model randomly selects a distance that is less than the Trip-Distance to be travelled in the first 'leg'. In this version of the model, the distances are chosen with a uniform probability distribution over the length of the Trip_Distance. To prevent Leg_Distances that are too short, the minimum Leg_Distance is 100 km, and thus a shorter Leg_Distance cannot be selected. If the Leg_Distance of the final leg is under 100km (a result of Markov-Chain), then the distance is appended to the previous leg thus no transfer to another mode occurs. Step 2 then randomly selects a mode of transport such as a Cargo Vessel, Truck or modified version such as Autonomous_Cargo_Vessel.

Step 3 calculates the Leg_Cost using equation 2:

$$Leg\_Cost_i = Leg\_Distance_i * Freight_{Weight_i} * Operational_{Cost_{Mode}} + Freight_{Weight_i} * Loading\_Cost \quad (2)$$

Where Freight_Weight is the user-selected weight of the shipment in metric tonnes selected in Step 1, Leg_Distance$_i$ is the randomly selected length of the leg of the journey, Loading Cost is the set value of loading for an inter-modal transport terminal in ($/Metric Tonnes) and Operational_Cost$_{Mode}$ is the measure of performance of the cargo transport mode randomly selected in Step 2 and is measured in $/(km*tonne). To capture the considerable variation in the performance within a given transport mode at a certain point in time, the variable Operational_Cost$_{Mode}$ is randomly selected from a log-normal probability distribution with given parameters µ and σ that are based upon the mean and standard deviation of the performance for that particular mode. The choice of a log-normal probability distribution is derived from the distribution of freight rates in Jacks and Pendakur (2010) that show the distribution of global shipping rates with a heavily-centred mean, a long, one-sided tail and no negative value for shipping rates. Equation 3 gives the calculations for the log normal parameters µ and σ² where m is the mean, and v is the variance.

$$\mu = \ln\left(\frac{m}{\sqrt{1+\frac{v}{m^2}}}\right), \quad \sigma^2 = \ln\left(1 + \frac{v}{m^2}\right) \quad (3)$$

Step 4 repeats step 2 and 3 to complete the length of the trip, randomly selecting a Leg_Distance and Transport_Mode until the sum of all Leg_Distances is equal to Trip_Distance. Note that this happens more easily due to the minimum trip distance of 100km. Each of the Leg_Costs are calculated using the process in Step 3, each time sampling off of a distribution of Operational_Cost$_{Mode}$ to find the Leg_Cost$_i$. Finally, Trip_Cost is calculated by the sum of all of the Leg_Costs.

Step 5 incorporates the future-oriented analysis into the model. Each Transport_Mode also has an attribute of Tech_Improvement_Rate that is based upon either historical data or other analysis as will be described in the following section. This Tech_Improvement_Rate is measured in %/year and is the representation of the generalized Moore's law for each technological domain. This measure is also represented by a log-normal distribution, based on the yearly distribution of improvement rates in Benson (2014) where the mean value is represented by the Technological Improvement Rate. The Operational_Cost attribute for each transport mode is then changed by the randomly selected Tech_Improvement_Rate for that year, giving an improved performance for each transport mode.

Step 6 is the Monte-Carlo aspect of the model, where many iterations are run for each year until the final year, thus giving a distribution of likely range of future outcomes that have incorporated the inherent uncertainty in the current and future performance of each of the transport modes.

### *3.2. Intermodal Transport Model Input Values*

This section derives the input parameters for the model that include both system-wide constants as well as transport mode attrivutes. The system-wide constants are constant throughout the entire model and are applied to all transport modes equally and include the following:

System Constants
- Handlings Costs – Mean
- Handling Costs - Stdev

The handling costs for an intermodal transport hub are taken from Janic (2007) as €2.8 per tonne and are adjusted to 2007 USD using the Purchasing Power Parity(PPP) (OECD, 2018). This is then converted to 2017 USD/tonne using the Consumer Purchasing Index (CPI) (BLS, 2018) giving a final value of $4.59/tonne. As an admittedly simple assumption in this model, a standard deviation of 25% of the mean is used for handling costs.

The Transport_Mode attributes are different for each transport mode and are used in the model for each leg based upon the Transport_Mode selected in step 2 as shown in Figure 1 above.

Each Transport_Mode has four main attributes:
- Operational Cost – Mean, Stdev
- Improvement Rate – Mean, Stdev

The mean values for the Operational Costs per mode are shown in 2017 USD per tonne-kilometer in Table 2 below.

Table 2: Current Operational Costs

| Domain | Reference Value | Measure | Year | 2017 Value | Source |
|---|---|---|---|---|---|
| Air Cargo | $1.766 | 2017$/tonne-km | 2016 | $1.669 | Bureau of Transport Statistics, 2018 |
| Ocean | $0.027 | 2017$/tonne-km | 2004 | $0.0196 | Bureau of Transport Statistics, 2018 |
| Truck | $0.241 | 2017$/tonne-km | 2007 | $0.227 | Bureau of Transport Statistics, 2018 |
| Rail | $0.058 | 2017$/tonne-km | 2005 | $0.046 | Bureau of Transport Statistics, 2018 |
| Inland Waterway | $0.0287 | 2017$/tonne-km | 2007 | $0.094 | PLANCO and BFG, 2007 |

Due to limitations in the data, the values in table 2 were calculated from historical values and then adjusted using the technological improvement rates from table 1 to create a 'current' value of transport mode performance. For example, the Air Cargo cost of $1.766/tonne-mile in 2016 will improve (decrease) at a rate of 5.5% per year to end at $1.669 in 2017.

### *3.3. Estimating Current Costs of Autonomous Transport Modes*

The current costs for autonomous transport modes have to account for the higher capital costs incurred due to automation. With the new modes of transport (i.e. autonomous modes) there is by definition significantly less publicly available data. Thus, the current operational costs are calculated using increase factors derived by comparing individual data points for a transport mode and their autonomous counterpart. Table 3 shows the estimated operational costs of the autonomous transport modes and the corresponding increase factors on which they are based.

Table 3: Calculated Operational Costs of Autonomous Transport Modes

| Domain | Estimated Current Autonomous Operational Cost ($/ton*km) | Current Factor for increased Autonomous Cost relative to non-autonomous transport mode | Source |
|---|---|---|---|
| Autonomous Air | $6.767 | 4 | Wheeler, 2012 |
| Autonomous Ocean Ship | $0.0249 | 1.26 | Kretschmann, 2015 |
| Autonomous Truck | $0.454 | 2 | Litman, 2018 |
| Autonomous Rail | $0.0828 | 1.8 | Mrazik Et al, 2015; Upton 2016; TEMS et al, 2010 |
| Autonomous Inland Waterway | $0.01197 | 1.26 | Kretschmann, 2015 |

The calculation for the initial Operational_Cost for autonomous Ocean_Ships and Inland Waterway Vessels, uses the analysis done by Kretschmann (2015) for the MUNIN project, which estimates lifecycle costs for different scenarios, the scenario chosen for the 2017 reference was the 'Reduced Crew Only' scenario except this analysis did not subtract the cost of the crew as the ships today still have crew on board even if they have all of the autonomous capabilities. In this scenario, the additional costs come out to $9.1 million, on top of a base cost of $34 million, resulting in an increase to the lifecycle cost of 26%. Thus, the assumed reference value for 2017 for an autonomous ocean or IWT ship is 26% higher than that of a standard ship ($0.0196 and $0.094), giving a reference value of the autonomous maritime ships as $0.0249 and $0.01197 as shown in Table 3.

The analog for autonomous trucks is the oft-studied autonomous car; Litman (2018) shows the average lifecycle cost of an autonomous vehicle to currently be approximately twice as expensive as that of a human drive vehicle. Thus the assumed autonomous truck cost is $0.454 per tonne-km as shown in Table 3.

The estimate for autonomous aircraft is given by Wheeler (2012) comparing the US military's MQ-9 reaper operating unit cost to similar aircraft (A-10 and F-16) of a cost of approximately 4 times as much currently. This gives an estimated current cost for aircraft of $6.676 per tonne*km for an autonomous cargo aircraft.

For autonomous rail, the comparison between a regular locomotive cost of $4 million (Mrazik et al, 2015) and the average cost from the Rio-Tinto autonomous freight train experiment in Australia of $9.77 million per train (Upton, 2016) gives a difference in capital cost of 244%, which accounts for ~33% of the total lifecycle cost (TEMS et al, 2010) giving an estimated increase of 80% above the non-autonomous rail cost.

### *3.4. Technological Improvement Rates for Autonomous Transport*

The expected annual technological improvement rate of 4.9% per year for the more complete metric of autonomous driving operational costs ($/mile) (Johnston and Walker, 2017; Bansel and Kockelman, 2017). This in-depth case took into account many aspects of the lifecycle cost including manpower, fleet management, fuel, depreciation, maintenance etc. and thus serves as a suitable reference for the expected improvement rate of autonomous transportation. As a baseline, this comparison is extended to the other forms of autonomous transportation and assume that the difference between the improvement rate of the non-autonomous mode and the autonomous mode are constant. Thus an improvement rate of 4.9% for autonomous driving gives an estimated difference of (4.9%-0.6% = 4.3%) for the additional autonomous capabilities. Thus, the estimated technological improvement rates for the autonomous transport modes are shown in Table 4 below:

Table 4: Estimated Technological Improvement Rates for Autonomous Transport Modes

| *Domain* | *Improvement Rate* |
|---|---|
| Inland Water Transport | 4.7% |
| Trucking | 4.9% |
| Rail | 6.2% |
| Ocean | 6.4% |
| Air Transport | 9.8% |

An attempt is made to capture some of the uncertainty provided by the assumption of a constant difference between the improvement rates of the autonomous and non-autonomous modes of transport by the sampling of the improvement rates on a log-normal distribution. Prior analysis of uncertainty in technological improvement rates has shown that a standard deviation of 50% of the mean is consistent with empirical data and can be used as a reasonable assumption (Benson et al, 2018; Triulzi, 2017).

### 4. Results – Distribution of Transport Mode Performance Over Time

Using the data from section 3 input into the model from section 2, simulations of future distributions of transport mode performance are run using a trip distance of 10,000km and a freight_weight of 50,000 tons and an end_year of 2050 with 1000 iterations per year. Analysing different trip distances and freight weights provide ample opportunity for future research.

## 4.1. Scenario 1 – All Autonomous and Non-Autonomous Modes

The first scenario shows the distribution of future costs for a multi-modal transport system that includes both autonomous and non-autonomous transport modes as is shown in Figure 3.

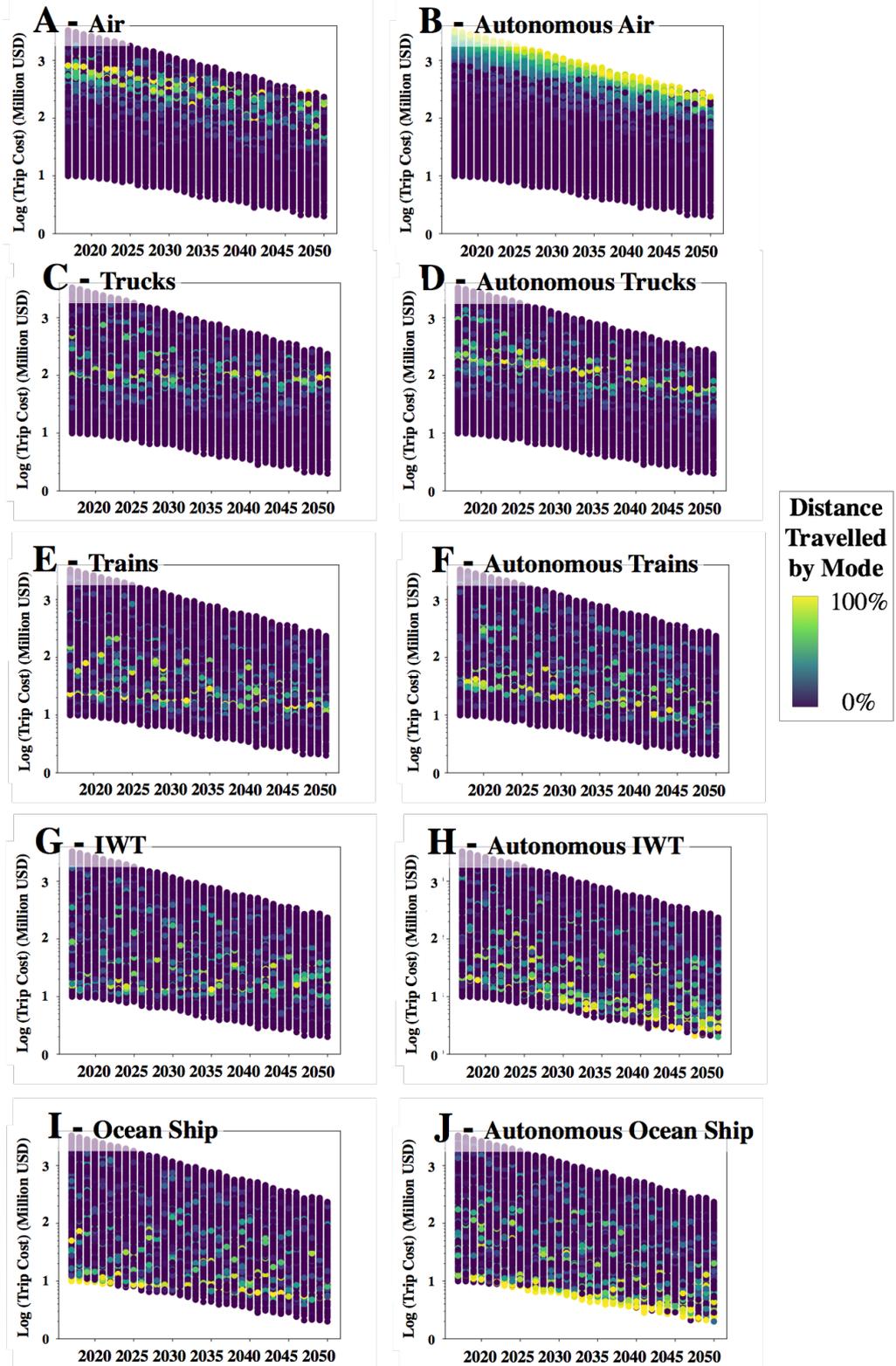

Figure 3: Future Distribution of Cargo Costs for Autonomous and Autonomous Transport Modes

The vertical axis for each chart is the total trip cost (measured on a logarithmic scale in millions of dollars), thus 1 on the vertical axis represents a trip cost of $10 million and 3 represents a trip cost of $1 billion. The horizontal axis shows discrete steps for years. The colouring of the data points represents the percent of distance that was travelled using the specified transport mode (i.e. Air Cargo in Figure 3(A)). Thus the yellow/green-coloured bands through the charts indicate a general trend of the cost competitiveness of each mode of transport as all the technologies change over time.

Figure 3 indicates that even through significant technological change over the next 30+ years, the relative cost competitiveness of the modes of transports is likely to stay the same. Figure 3(A) and (B) show that both the Air and Autonomous Air modes are always among the most expensive modes of transport and Figures 3(I) and (J) show that both the Ocean Ship modes are always among the least expensive modes of transport. The remaining modes of transport stay in the middle and are largely competitive with each other as would be expected.

Another notable point is that the spread between the highest and lowest costs becomes much tighter as the most expensive modes, Air Cargo and Autonomous Air Cargo, improve a relatively high rate compared to the other domains. For example, the high end of the total trip cost in 2018 in Figure 3 show costs higher than $3.5 Billion for the shipment, with the highest cost decreasing to slightly above $200,000,000 by 2050. These costs are likely much higher than the market will bear, leading to many shipments of this size and weight using ocean cargo for years to come, even given the relatively rapid improvement rate of the Air Cargo transport mode. This is contrasted with the lowest costs decreasing from around $10 million in 2018 to approximately $2 million in 2015. This indicates that while the distribution of the costs may not change significantly, the difference between the lowest and highest cost modes will narrow.

### 4.2. Scenario 2 – Autonomous vs Non-Autonomous Specific Modes

The second scenario, as shown in Figure 4, explores intermodal transport when there are only the autonomous and non-autonomous versions of one type of transport (i.e. Autonomous Trucks vs Conventional Trucks). The axes are the same as in Figure 3, with the most notable trend being the changing distribution of costs between the yellow (Autonomous) 'band' and the purple 'Conventional' bands.

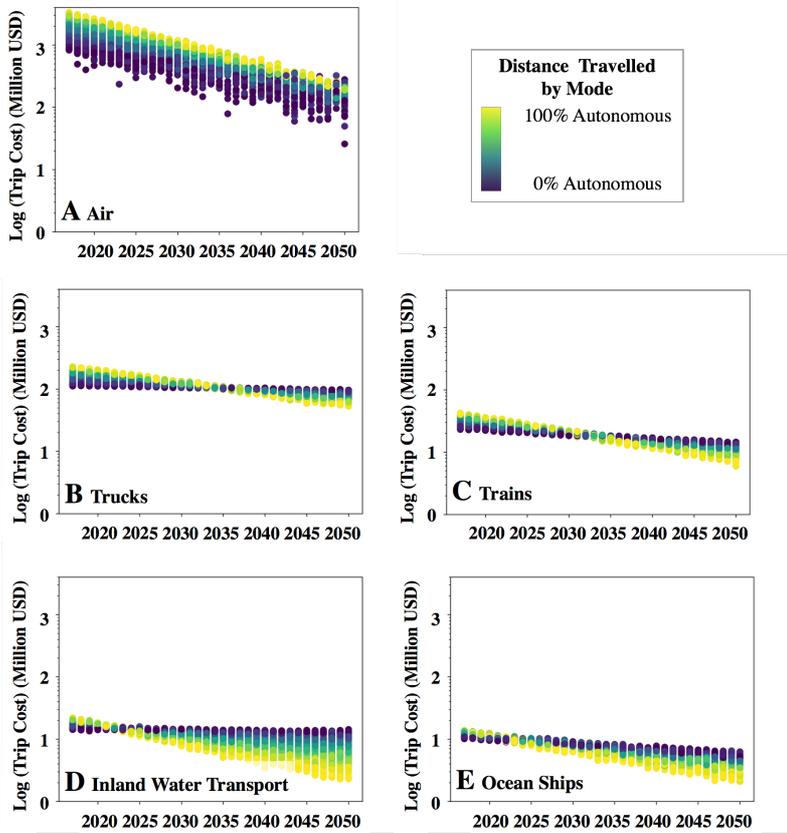

Figure 4: Future Distribution of Cargo Costs for Autonomous vs Non-Autonomous Single Mode Transport

In Figures 4(B) through (E) it is possible to see the 'cross-over' time ranges when the cost of the autonomous and non-autonomous mode of transport are competitive when the yellow 'band' (highly autonomous trips) crosses the 'purple' band (highly non-autonomous trips). Figure 4(A) does not show a complete cross over time range, but rather shows that in the end of the time range (2050) Air Cargo and Autonomous Air Cargo are just starting to be competitive with each other. This is due to the fact that non-autonomous Air Cargo is already improving at a relatively high rate and that the current difference in cost for the addition of autonomy is the highest at four times the cost of the non-autonomous mode as shown in Table 3.

Autonomous Cargo Ships are shown in Figure 4(E) compared to their non-autonomous counterparts. The results show a stark difference in improvement between the two modes, with the autonomous mode quickly becoming competitive with its conventional counterpart in the mid-2020s and offering significantly cheaper rates later into the middle of the 21$^{st}$ century. One interesting point to note in this chart is that the density of the change is not as high as in the Air Cargo comparison, with a wider band of high autonomous ocean shipping (noted in yellow) and non-autonomous (noted in purple) in the later years. This would indicate that the autonomous and non-autonomous domains will likely be more competitive in the future and thus are more likely to co-exist for longer periods. This differs from some technology transitions in fast moving domains where the arrival of an improved technology all but eliminates the presence of the replaced technology (i.e. internet audio vs cds).

The transition between inland water way vessels as shown in Figure 4(D) is estimated as the mid-to-late 2020s, with the early date once again due to the relatively low estimated cost difference between an autonomous and non-autonomous IWT vessel. As time progresses, autonomous IWT vessels maintain tight bands, but a wide spread between the autonomous and non-autonomous modes. This indicates autonomous IWT vessels gaining a clear cost advantage over the non-autonomous counterpart.

Both ocean shipping and inland water transport vessels show relatively soon 'cross-over' dates, which is certainly due to the baseline assumption that autonomous shipping is currently only ~1.26 times as expensive as non-autonomous ocean shipping. A higher assumed current cost difference or a lower assumed technological improvement rate difference would certainly shift the cross-over time. Table 5 shows the sensitivity of the 'cross-over' date to different assumptions.

Table 5: Current Cost Increase Multiple vs ΔTechnological Improvement Rate%

|    | 1.5X | 2.5X | 3.5X | 4.5X |
|----|------|------|------|------|
| 2% | 2039 | 2067 | 2086 | 2098 |
| 4% | **2028** | 2043 | 2052 | 2059 |
| 6% | 2025 | 2035 | 2041 | 2045 |
| 8% | 2024 | 2031 | 2036 | 2037 |

As is shown in Table 5, the baseline assumptions (1.26X, 4.3%) from the literature for ocean and inland water way transport vessels provide a cross-over data consistent with the qualitative determination from Rolls Royce (2016) of 2025, but this number can be pushed out considerable to later years given different assumptions, and thus it would not be surprising given the uncertainty in the data to have a cross-over date in the 2030s or early 2040s for autonomous maritime technologies or even later, aligning with the expectations of an industry leader that autonomous ocean ships will not be cost-competitive in 'his lifetime' (Wienberg, 2018)

5. Conclusions

There has been much discussion about how autonomous will affect the future of shipping, this paper uses quantitative, probabilistic modelling to gain a better understanding of how that future may look. The results show that gains in autonomy are likely to provide significant decreases in costs, but will be more impactful for the modes of transport that are currently more expensive, such as Air Cargo transport. Our analysis indicates that if autonomy affects all modes equally then the relative distribution of cargo costs is not expected to change significantly, however the absolute difference between the lowest and highest cost modes will be dramatically reduced. That is to say that autonomy will accelerate the shrinking of the cost difference between ocean and air cargo shipping. For all of the modes except ocean shipping, the introduction of Autonomy results in a relatively stark transition whereby the autonomous mode quickly and permanently gains a significant cost advantage over its non-autonomous counterpart. Finally, the lowest cost modes such as ocean and IWT shipping are likely to maintain healthy cost competitiveness with autonomous version of the other modes of transportation into the middle of the 21st century even without adopting the autonomous modes themselves.

This analysis may indicate that due to the significant cost advantage currently held by ocean shipping that a 'wait-and-see' approach to autonomous technologies in other modes may be a reasonable strategy as there is not as much risk to rapid disruption in the ocean shipping industry as there is in the other transport modes.

One final takeaway is that the significant amount of uncertainty inherent in the data and in the forecasts indicates a need for better data collection, monitoring, analysis and research into how autonomy will impact the cargo transport industry through accurate collection of current cost data of autonomous transport modes as well as their technological improvement rates. Doing so will significantly decrease the uncertainty around the timing and effects of the transition of autonomous transport modes.

## 6. Acknowledgements

The author would like to thank TU Delft, the National Netherlands Defence Academy and the US Air Force for contributing to this research. The views expressed in this paper are that of the authors and do not necessarily represent the views and opinions of the US Air Force.

This article was created with support from NOVIMAR Project, that is concerned with creating a NOVel Iwt and MARitime transport concepts. This project is supported by the European Commission under Grant 72300.